# Application of the density functional theory to the fuel cell problem

Sergey Stolbov, Marisol Alcantara Ortigoza and Talat S. Rahman

*Department of Physics, University of Central Florida, Orlando, Florida 32816, USA*

*Abstract:* The large-scale practical application of fuel cells in hydrogen economy is possible only with a dramatic reduction of the cost coupled with a significant improvement of the electrocatalytic properties of the electrodes. This goal can be achieved through a rational design of new materials, which requires understanding of the microscopic mechanisms underlying the electrocatalysis. We discuss some applications of the density functional theory to this problem using using alloying of Ru with Pt as the case in point. We provide some details for Pt islets on Ru nanoparticles, for which our calculations explain the high CO tolerance observed for such anodes of the fuel cell. We also discuss energetics responsible for the stability of these islets on the nanoparticles.

## 1. Introduction

Fuel cells, invented more than 150 years ago, are currently of great interest, because they are a key element of the emerging hydrogen economy. One of the most promising types of fuel cells is the Proton Exchange Membrane Fuel Cell (PEMFC), because it can be used for powering of transportation vehicles, which is a major application of the hydrogen economy [1]. In PEMFC, protons, released in the course of hydrogen oxidation at the anode, move through a solid polymer electrolyte membrane to the cathode. At the cathode, the protons meet supplied oxygen and electrons transferred from the anode through the circuit, which results in the electrocatalytic oxygen reduction reaction (ORR) with water as the final product. PEMFC is a clean source of electric power with low operating temperature (60 – 80 $^o$C) and high power density. In contrast to PEMFC, in Direct Methanol Fuel Cells (DMFC's) hydrogen is not supplied from a separate source, but is produced from methanol on the same anode that is used for the hydrogen oxidation. This device provides power with lower density than PEMFC. However, its advantage is the fuel, liquid methanol, which is easy to store. DMFC can be used as the electric power source for smaller devices, such as cell phones, laptops, etc.

These fuel cells thus offer great advantages for various applications, although several obstacles need to be overcome before their large scale implementation. A major hindrance is that fuel cells are unacceptably expensive. To make them economically competitive with conventional technologies, their cost has to be lowered by a factor of 10 [1]. Since the Pt-based catalysts, used in both electrodes of the fuel cell, make up a major part of the cost, search for new electrocatalytic materials with a reduced loading of precious metals is critical for commercialization of PEMFC and DMFC.

There are also significant shortcomings in the functionalities of the electrocatalysts. PEMFC usually operates with hydrogen obtained from hydrocarbons, which inevitably contains carbon monoxide. Even small traces of CO, remaining in the gas after purification, poison the Pt anode by blocking its active site and this way suppressing hydrogen oxidation [2]. In DMFC, CO released with the methanol oxidation, is supposed to be oxidized with OH obtained from admixed water. Nevertheless, it severely poisons the Pt anode. The PEMFC performance suffers even more from a low rate of ORR on the Pt cathode, which results in a high overpotential, and hence low fuel cell efficiency [3]. In DMFC, methanol, undesirably transferred onto the Pt cathode, additionally reduces the ORR rate [4].

Clearly, the great advantages of fuel cells can be efficiently utilized only if the cost of the electrodes is dramatically reduced and their electrocatalytic properties are significantly improved. It is thus not surprising that enormous effort is made to find new materials, which meet these requirements. Noticeable progress has already been made in improving the anode CO tolerance. It is known that alloying of Pt with a second (and even third) metal element may reduce this poisoning effect. For instance, Pt-Ru alloys are found to be more tolerant to CO than pure Pt [2]. Alloying Pt with Sn [2] and Mo [5] also improves anode performance. However, these anodes are still strongly affected by CO poisoning. Another disadvantage of these materials is high loading of expensive platinum. Not surprisingly the report that nanoclusters of Ru with sub-monolayer of Pt (PtRu$_{20}$) are much more tolerant to CO poisoning than commercial PtRu catalysts [6,7] has been welcomed with optimism. It is also important that the content of Pt in these novel materials is much lower than that in Pt-Ru alloys. From estimate of the average diameter of Ru nanoparticles (2.5 nm) and Pt/Ru ratio, the authors conclude that the deposited Pt forms small islands (islets) on the facets of the Ru nanoparticles. Some progress has also been achieved in improving the cathode properties. Apart from reduced Pt loading, some of the PtM alloys (M = Ti, V, Cr, Mn, Fe, Co, and Ni) [8, 9], and Pt mono-layers on the Pd(111) surface [10] are found to demonstrate the improved ORR kinetics.

Clearly, the reactivity of catalysts is determined by the nature of chemical bonding between the catalyst and reactants or intermediates, which in turn depends on the composition, geometric structure and size of the catalyst particles. Few nanometer size particles usually have an increased number of steps at the surface and chemically undersaturated sites (CUS) at the facet edges. This is important, because CUS are naturally more reactive and the steps at metal surfaces are known to dramatically enhance oxygen dissociation [11], one of the key elements of ORR. Further decrease in nano-particle size may bring to play the quantum size effects, which can substantially change catalytic properties. Search for efficient catalysts can thus be performed by varying composition and particle size and by modification of surface structure of the system. Since this involves many variables, the search can be efficient if there is a prior understanding of the microscopic mechanisms controlling the catalytic processes.

In heterogeneous catalysis these mechanisms essentially depend on the energetics and pathways of the elementary steps (adsorption energies and diffusion and reaction activation barriers for reactants, intermediates and products on the catalyst surface) are very important microscopic characteristics, which can be obtained from first principles calculations based on the density functional theory (DFT) [12,13] and then used to estimate the rates of the elementary processes and determine the rate limiting steps [14]. Analysis of the electronic charge densities and densities of electronic states calculated for various stages of the reaction can be used to reveal the factors controlling the reaction energetic and hence make a basis for rational design of new catalysts. Finally, the activation barriers obtained from DFT calculations can be used in the kinetic Monte Carlo (KMC) simulations to obtain the temperature and pressure dependencies of the total reaction rate [15]. This is an excellent example of the multiscale approach, in which the microscopic characteristics, obtained within DFT for zero temperature, are used to describe the mesoscopic kinetic processes.

The electrocatalytic kinetics in the fuel cells is a more complicated phenomenon, which involves also the electron and proton transfers between two electrodes with different Fermi-levels. However, as found for heterogeneous catalysis, in this case too DFT-based calculations may provide key microscopic characteristics, which are used as parameters for kinetic models. Such techniques are nicely described in a review article by Shi and co-authors [16]. Of the various approaches available, we would like to mention here a simple and elegant one proposed by the Nørskov and co-authors [17], in which they first calculate adsorption energies of reactants and intermediates using a DFT-based method. If some of the adsorption configurations are achieved through the process involving the electron transfer, relevant bias effects are taken into account by shifting the energy by $-eU$, where $U$ is the electrode potential. In this manner a diagram of the free energies of the configurations, as a function of the electrode potential, is created. Although this method results not in kinetic, but in thermodynamic characteristics, it can be very useful for understanding the electrocatalytic processes. For instance, using this approach Norskov et al. [17] have revealed the origin of the overpotential for ORR on the Pt cathode.

In electrocatalysis, some important processes (e.g. $O_2$ dissociative adsorption on the fuel cell cathode, CO adsorption and diffusion on the anode) do not involve the electron or proton transfer which make them similar to the elementary processes of conventional heterogeneous catalysis. The only difference is that the reactants experience the electric field caused by the potential difference between the electrodes. This effect has been estimated for the O and OH intermediates adsorbed on Pt [17]. Assuming the dipole moments of adsorbed O and OH to couple with the electric field caused by change in the potential within the 3Å electrochemical double layer, Norskov et al. found the energy of such interaction to be negligible. This means that the elementary processes, which do not involve the electron or proton transfer in fuel cell, may be characterized using methods of conventional heterogeneous catalysis. This approach is extensively applied to the fuel cell related problems. The density functional theory has long been used as an insightful and reliable theoretical and computational tool in understanding the fundamental steps in heterogeneous catalysis. Its application to fuel cells is not surprising. While a full review of the subject is beyond the scope of this paper, we present here a short discussion of its application to unravel the mechanism of CO tolerance of Pt/Ru nanostructures reported in Ref. [6], and beyond that in analyzing the stability of these nanostructures.

## 2. First principles studies of CO energetic on Pt-Ru nanostructures: rationale for tolerance

CO poisons Pt anode reactivity by blocking its active sites, and, therefore, the obvious way to reduce the poison effect is to remove of CO from these sites. Since the removal is supposed to depend on the strength of the molecule-surface binding, a number of publications have been focused on the mechanisms of the CO adsorption on Pt-Ru systems [18-20]. For example, Koper et al. [18] have calculated from first principles CO adsorption energy $E_{ad}(CO)$ on clean Pt(111) and Ru(0001) surfaces, as well as on a Pt monolayer on Ru(0001) and a Ru monolayer on Pt(111). Since $E_{ad}(CO)$ is found to be the lowest for the case of a monolayer of Pt on Ru(0001) ($Pt_{ML}$/Ru(0001)), Koper et al. propose it to be the rationale for high CO tolerance for this system. First principles study of alloying effects on CO adsorption on Pt [19,20] suggest that strain induced by the second element modifies the electronic states of Pt in such a way that it causes a decrease in CO adsorption energy. Subsequent studies [21,22] have also upheld the view that enhanced Pt tolerance to CO poisoning is associated with a decrease in $E_{ad}(CO)$.

In general the removal of the molecule from the Pt sites can be achieved by its desorption or diffusion. However, in the above mentioned case of $Pt_{ML}$/Ru(0001), in which only Pt atoms are exposed to the surface, the only mechanism for CO removal is desorption. Desorption rate $R$ can be estimated using the transition state theory:

$$R = D_0 e^{-\frac{\Delta E}{kT}} \quad (1)$$

Where $D_0$ is the pre-factor, and, for this particular process, $\Delta E = E_{ad}(CO)$. Setting $D_0 = 10^{12}$ $sec^{-1}$, which is a typical value for the pre-factor, $T = 350K$ (operation temperature for PEMFC), and taking $\Delta E = 1.11$ eV from Ref. 18, we obtain R ≈ $10^{-4}$ $sec^{-1}$. The desorption rate is thus very low for $Pt_{ML}$/Ru(0001), and it is expected to be even lower for other anodes, because of higher $E_{ad}(CO)$.

The observed enhancement of the CO tolerance of the Pt/Ru nanostructure cannot thus be caused by increase in the desorption rate. Missing from this analysis is the consideration of CO diffusion rates which for anodes with inhomogeneous surfaces such as the $PtRu_{20}$ nanoparticles [6], may be the main factor contributing to the CO removal from Pt sites. Spillover of CO from Pt islands to the Ru substrate is mentioned in Ref. [21] as a possible mechanism for the high CO tolerance, but the argument contrives to be based on the assumption of weakened CO adsorption. The spillover rate, however, depends on the diffusion of the molecules, which is implicitly related to its adsorption. Conclusion about efficiency of the CO spillover needs thus to be based on knowledge of activation energy barriers for CO diffusion in the system, which can be obtained from accurate DFT calculations. We have performed such calculations for the CO diffusion on system modeling the the $PtRu_{20}$ nanoparticles [6], which we take to be Pt islands on the Ru(0001) surface (presumably the dominant facet of the particle). We have calculated the energetic of the system along the entire path of the CO molecule diffusing from the center of the Pt island to its edge and further to the Ru substrate.

## 2.1. Computational Details

Our calculations have been carried out within DFT using the plane wave pseudopotential method [23] as embodied in the code VASP [24] with ultrasoft pseudopotentials [25]. To maintain periodicity of the systems we used supercell comprising of a 5 layer Ru(0001) slab with a four or seven Pt-atom island, on one side, and vacuum layer of 15 Å. Most calculations also included the CO molecule adsorbed either on the island or on the Ru(0001) substrate. To diminish interaction between the periodic images of Pt islands, the supercell was extended along the (0001) surface making up the (4x4) superstructure. With such geometry the shortest distance between edges of neighboring islands equaled two Ru-Ru bond lengths. The supercell contained 80 Ru atoms, plus Pt atoms forming the island and a CO molecule. The Brillouin zones were sampled with the (3x3x1) Monkhorst-Pack $k$-point meshes [26]. We used a kinetic energy cutoff of 400 eV for the wave functions and 700 eV for the charge density which provided sufficient computational accuracy for the oxygen containing structures. The Perdew-Wang generalized gradient approximation (GGA) [27] has been used for the exchange-correlation functional.

To achieve structural relaxation, a self-consistent electronic structure calculation was followed by calculation of the forces acting on each atom. Based on this information the atomic positions were optimized to obtain equilibrium geometric structures in which forces acting on atoms do not exceed 0.02 eV/Å.

## 2.2. Results and Discussion

We present here results for the 7-atom Pt islet on the Ru(0001) surface (7Pt/Ru(0001)). We have calculated the energies of CO adsorption on the top of the central (c-Pt) and edge (e-Pt) platinum atoms and on two non-equivalent (fcc and hcp) hollow sites, as well as on the Ru substrate site neighboring the Pt islet (n-Ru) and the next neighbor Ru site (nn-Ru). In addition, activation energy barriers have been calculated for CO diffusion from c-Pt to e-Pt, between two e-Pt (along the island edge), from e-Pt to n-Ru, and from n-Ru to nn-Ru sites. Fig. 1 shows the energetics calculated for the system with CO moving along the c-Pt – e-Pt – n-Ru – nn-Ru path.

As illustrated in Fig. 1, CO bonding increases, as the molecule moves from the center of Pt island to its edge and further to the Ru(0001) substrate. This negative gradient of $E_{ad}(CO)$ along the c-Pt – e-Pt – n-Ru – nn-Ru path already indicates that CO molecule adsorbed on 7Pt/Ru(0001) would prefer to leave the islet for the Ru substrate. Furthermore, we find $E_{ad}(CO)$ = -1.94 eV for CO on clean Ru(0001). This value is lower than those for CO adsorption on the n-Ru and nn-Ru sites suggesting that CO tends not only to leave the Pt islet, but also to move away from it.

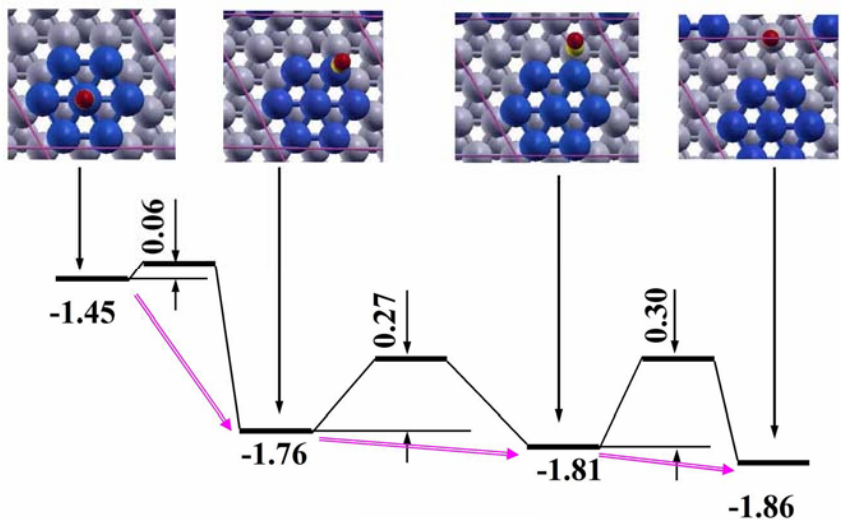

Fig.1.

The activation energy barrier for CO diffusion from c-Pt to e-Pt through the bridge site is found to be as low as 0.06 eV. The barriers for the rest of the considered path are also quite low. The highest n-Ru – nn-Ru barrier is 0.3 eV resulting in a diffusion rate $R \approx 5*10^7 \ sec^{-1}$ from Eq.1 with $D_0 = 10^{12} \ sec^{-1}$ and $T = 350K$. Clearly, this rate is much higher than that for CO desorption. We thus find the spillover of CO from Pt islets to Ru substrate to be a favorable process, which keeps active Pt sites available for hydrogen oxidation and hence provide the high CO tolerance of the $PtRu_{20}$ nanostructure. Interestingly, we find CO bonding to the Pt island atoms to be significantly stronger than its bonding to $Pt_{ML}/Ru(0001)$ (we obtain $E_{ad}(CO)$ = -1.15 for $Pt_{ML}/Ru(0001)$, which differs slightly from -1.08 eV reported in Ref. 6). For the edge Pt atom it is even stronger than CO bonding to Pt(111) (-1.6 eV). Nevertheless, this system provides very efficient mechanism for CO removal from active Pt sites, and this mechanism originates not from weak CO bonding to Pt atoms, but from the negative adsorption energy gradient and low energy barriers for CO moving from the center of Pt island to its edge and further to the Ru(0001) substrate. The calculations have also been performed for a 4-atom Pt islet on Ru(0001), in which all Pt atoms belong to its edge. It is thus not surprising that the obtained $E_{ad}(CO)$ = -1.78 eV is almost the same as for the edge Pt atoms in 7Pt/Ru(0001) (-1.76 eV). We can thus expect the same spillover scenario for smaller islands.

## 3. First principles studies of the energetic of Pt islets on Ru surfaces/facets: rationale for nanoalloy stability

The above spillover mechanism of the CO removal from the catalytically active Pt sites of the Pt-Ru nanostructure appears to depend critically on the size of the Pt island. This naturally raises the question of stability of the structure with several atom Pt islands on the Ru substrate, which is related to the problem of the heteroepitaxial growth. Theoretical models of heteroepitaxial growth suggest that the growth mode is determined by the competition of factors such as the surface energies of the bare substrate and the heteroepitaxial layer, the interface free energy, and the strain energy introduced by the lattice mismatch of the two species [28-30]. For example, a mismatch in the respective bulk lattice constants strains the interface and may set off the 3D clustering growth mode [29]. On the contrary, growth of ad-layers with lower surface energy than the substrate may favor the 2D layer-by-layer growth mode [29]. Considerations along these lines, however, lead to ambiguities in predicting Pt growth on Ru(0001). While the higher cohesive energy of Ru (relative to Pt) may [31,32] imply that the surface free energy of Ru(0001) is higher than that of Pt(111) and point to 2D layer-by-layer growth, the stress caused by the Ru-Ru and Pt-Pt bond length misfit (the bond lengths in bulk Ru and Pt are 2.706Å and 2.775 Å, respectively [33]) may lead to 3D clustering. There may also be competition between the above two factors, leading to a critical Pt island size at which there is crossover between 2D and 3D growth mode or island-substrate atom exchange [29,34]. One of the goals in this work is to determine whether there is indeed a critical size beyond which 2D Pt islands are no longer stable on Ru(0001).

In the experiments in question [5,6], Ru nanoparticles have well developed facets divided by edges. Since one of the dominant facets has the Ru(0001) geometry [6], as a first step in the modeling of the Pt-Ru nanoparticles, we consider the formation of Pt islands as a function of size, on the Ru(0001) surface. To this end, we carry out first principles calculations of the system total energy to determine the geometry and formation energy of Pt islands, as well as that of 1ML of Pt on Ru(0001). Clearly, the main drawback of such calculations is that the predictions are relevant to zero temperature and samples relaxed in infinite time. The kinetic of the system can be partially included by taking into account diffusivity of Pt ad-atoms. At least for low temperatures, the three aforementioned growth modes can be understood in terms of diffusion barrier differences between on-step hopping and step-descending hopping (Schwoebel barriers) [35-37]. If the Schwoebel barrier is positive, the probability for the ad-atoms to be trapped on the step terraces is high, and 3D clustering is favored. On the contrary, if the Schwoebel barrier is zero or negative, the ad-atoms that happen to lie on the step terrace are more likely to hop onto the substrate and favor the 2D growth [36]. To understand the role of such effects in the formation of Pt islands on Ru(0001) we also calculate the barriers for diffusion of Pt atom from the 8 atom island to the Ru substrate.

### *3.1. Energetics of Pt islets on Ru(0001)*

To reveal stable configurations for the sub-monolayer Pt deposited on the Ru(0001) substrate, we have calculated the optimized geometric structure and energetics of the 1 to 9 Pt atom islands and one Pt monolayer on Ru(0001) using the (4 x 4) supercell. To characterize stability of a given Pt island, we obtain its formation energy, which is defined as:

$E_{form} = E(Ru+Pt) - E(Ru) - nE(Pt_{at})$.

Here, $E(Ru+Pt)$ is the total energy of a Ru slab adsorbed with a n-atom Pt island, while $E(Ru)$ and $E(Pt_{at})$ denote the total energies of the clean Ru slab and a free Pt atom, respectively. Note that the formation energy of stable structures is negative. The structure with lowest average formation energy per Pt atom, $E_{form}/n$, is thus distinguished as the energetically most favorable one. Fig. 2 presents $E_{form}/n$ as a function of the size of the island. Note that the horizontal line at the bottom of the panel marks the formation energy per atom for Pt monolayer on Ru(0001) which is found to be -5.89 eV. We find that the larger the Pt island is, the higher its stability. Such a trend culminates and is confirmed at full monolayer Pt coverage, which provides lowest $E_{form}/n$.

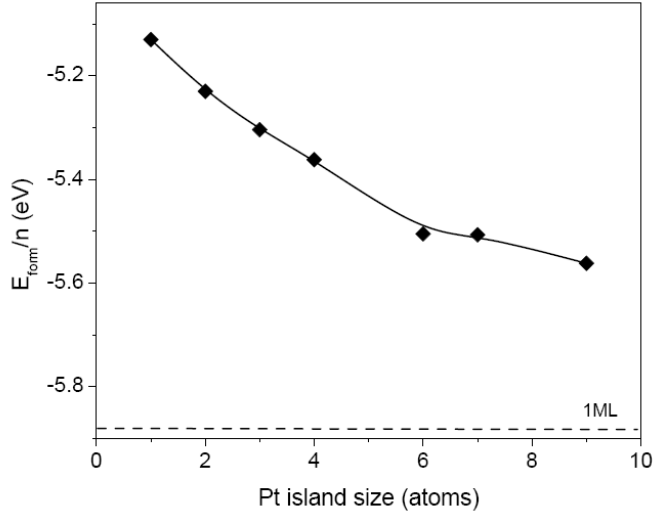

Fig. 2

The effect of the Pt atom detachment from islands has also been studied. Fig. 3 shows two configurations considered for the 7-atom Pt island adsorbed on Ru(0001). We find the detachment (transition from the left to the right configuration in the figure) to cause an increase in $E_{form}$ from -38.55 eV to -37.52 eV. Similar results have been obtained for the islands of other sizes.

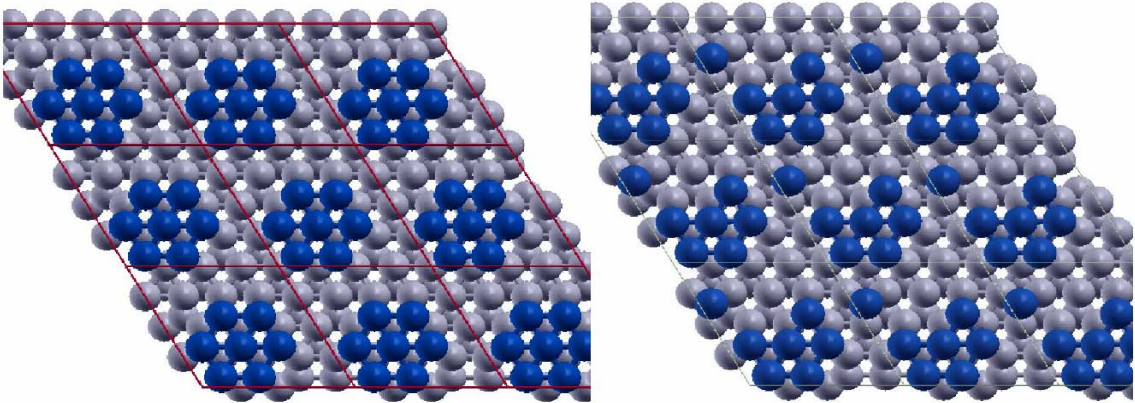

Fig.3

The increment of energy per Pt atom upon detachment, however, does not depend significantly on the island size and vary in the range of 0.11 - 0.14 eV. For instance, for a 2-atom Pt island, detachment leads to an increase in $E_{form}$ from -10.46 eV to -10.24 eV while, for a 3-atom island, the energy increases from -15.88 eV to -15.52 eV upon the detachment. We thus obtain a clear trend: the larger the two-dimensional Pt island is (up to 1 ML), the lower its formation energy per atom. Thus, assuming that the free energy of the system in consideration is dominated by its DFT total energy, we conclude that Pt tends to wet Ru(0001).

We have also performed calculations for some 3D Pt configurations on Ru(0001) and found that their 2D isomers have lower energy. For example, two configurations of a 9-atom Pt island with 2D and 3D structures (see Fig. 4) were found to have $E_{form}$ = -50.06 eV and -48.54 eV, respectively.

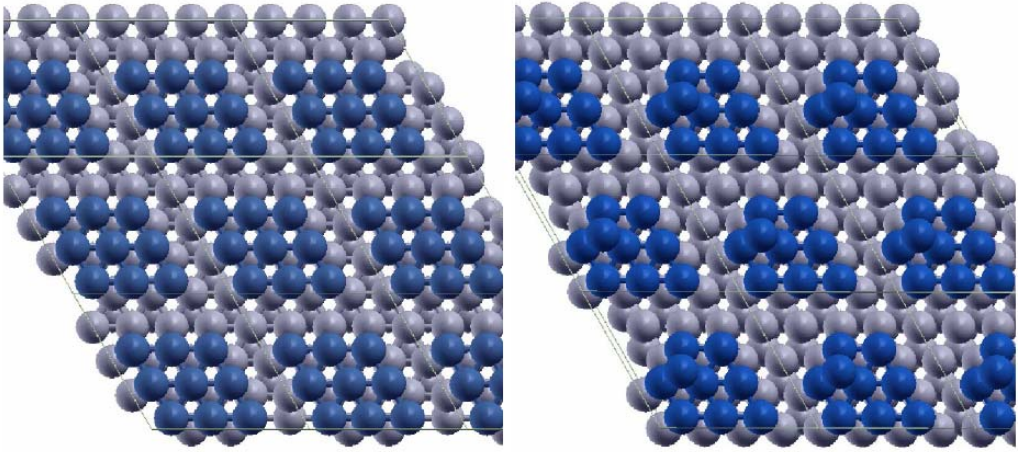

Fig.

The above results are indicative of the propensity of Pt atoms to increase their local coordination and thus form 2D islands up to full coverage when adsorbed on Ru(0001), which overpowers the stress effects derived from the Pt-Ru bond length misfit. This is justified only at low temperatures and after a long relaxation time. As mentioned above, to take into account the kinetics of the system, we have also looked at the diffusivity of a Pt ad-atom in the event that it is adsorbed on top of Pt islands. The question is thus whether the Pt ad-atom will be trapped on top of the Pt island (as shown in the right inside of Fig. 4) or will descend at the kink site on the Ru(0001) surface (left inside of Fig. 4). Our calculated activation barrier of the Pt monomer diffusion on the 8-atom Pt island from hcp to fcc is 0.23 eV and from fcc to hcp is 0.09 eV. The activation barrier for the descendant step of a Pt monomer from fcc to the kink site on the Ru(0001) is 0.39 eV (and 1.92 eV for the inverse process). We thus obtain the Schwoebel barrier of the heteroepitaxial descendant step of Pt to be +0.30 eV. Since it is positive, it points to 3D clustering growth. However, a word of caution must be inserted here. The Ru nanoparticles that we are trying to understand are decorated with Pt at very low coverage (~ 0.1 ML) through spontaneous deposition via a $[PtCl_6]^{2-}$ + 0.1 M $H_2SO_4$ solution [6]. With such technique, the probability of a Pt ad-atom to fall on top of the island is low. Besides, there is indication [37] that, even in the case of positive Schwoebel barriers, as long as the atom deposition rate is low and the diffusion speed fast, the reflective property of steps is valid only at temperatures well below room temperature for fcc(111) metals. In that regard, it is worth mentioning that, as reported [38], at low coverage of 0.03 ML of Pt, 2D islands (of ~ 2 nm) are formed on Ru(0001) and the 2D/3D crossover depends on experimental conditions.

There still remains the question: why, despite the misfit, 2D configurations are favorable? In order to grasp some understanding of the issue, we note first that in the 2D Pt islands under consideration, the number of NN varies from 3 (single atom) to 9 (full monolayer). Decrease in the number of NN usually causes reduction of the equilibrium interatomic distances. Indeed, we find that for a free standing Pt monolayer, in which every Pt atom has only 6 nearest neighbors, the equilibrium Pt-Pt NN distance is much shorter (~2.6 Å) than that in bulk Pt (~2.8 Å) and even shorter than the Ru-Ru NN distance in bulk Ru (~2.71Å). The misfit in low dimensional structures is thus not a well defined quantity because of the dependency on the coordination number of the atoms in question. For these surface alloys, there is also the issue of stress induced by the bond-length misfit between the Pt nanostructures and the Ru surface atoms, neither one of which is expected to be at the bulk value, given the diversity of their local geometric environment. For the Pt atoms on the top of hcp metal such as Ru, there is also incommensurability in bulk structure. We find that the bulk NN bond length of Pt atoms certainly decreases when they arrange in an hcp structure. In that case, the bulk bond length misfit between Pt and Ru decreases from 2.8% to 1.4 %. Furthermore, the surface interlayer distance in Pt(111) expands to 2.49Å (1.0 % with respect to bulk), while that of the hypothetical Pt(0001) contracts to 2.39 Å notwithstanding that intralayer NN distances are 1.8 % smaller than in the fcc bulk.

### 3.2. Modeling Pt diffusion through the edges of Ru nanoparticles

From the above, we have gained an understanding of the tendency of Pt atoms to form 2D layer wetting the Ru(0001) surface rather than clustering in multiple 2D or 3D structures. In this case, however, even for low coverage (~0.1 ML), a large island should totally cover one of the facets of the Ru nanoparticles. For example, for the Ru nanoparticle of 2.5 nm with the proposed cubo-octahedral structure [7], the 0.1 ML of Pt coalesced into a single island would totally cover one of the squared facets, while experiments suggest that Pt islands maintain 5 to 7 atom size on Ru nanoparticles for such coverage [7]. To resolve this discrepancy we need to note that one main difference between the infinite Ru(0001) surface and the Ru nanoparticle is that the latter exhibits edges dividing its facets. These edges may be obstacles for Pt atom diffusion. If this is true, 2D islands formed on each facet do not join together into a large unique island because the edges prevent those initial small islands from diffusing to other facets, thus persisting as few-atom 2D islands.

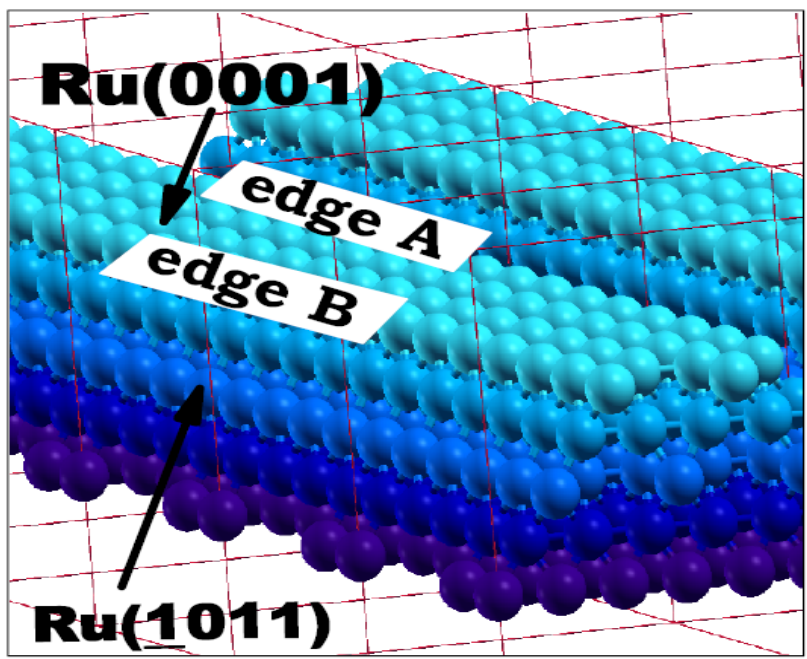

Fig. 5

To check this hypothesis, we simulate two kinds of edges of the Ru nanoparticle and calculate diffusion barriers of Pt monomers and dimers through these edges. We model the diffusion through the edges formed by facets of (0001) and (1101) geometry, which are among the most stable Ru surfaces [7,39]. The supercell in this case contains 116 Ru atoms and made of a 4-atom wide Ru(0001) facet and two Ru(1101) facets (see Fig. 5). The construction of this Ru supercell, which has 7 x 4 in-plane periodicity, is achieved by stacking five Ru(0001) layers: two of 7 x 4, one of 6 x 4, one of 5 x 4, and one of 4 x 4 atoms. The so obtained edges, on each side of the Ru(0001) have different local geometry which for convenience are labeled as A and B. Atoms forming edge A (edge B) are contiguous to hcp (fcc) hollow sites of the (0001) facet. The bottom two layers (see Fig. 5) were not allowed to relax to guarantee the stability of the superstructure. We impose a 15 Å vacuum layer between periodic superstructures along the direction perpendicular to the surface, as in the system described previously. The Brillouin zone is sampled with a (2x3x1) k-point mesh. The adsorption energy and diffusion barriers of Pt monomers and dimers are calculated on the (0001) and the (1101) facets.

*3.2.1. Pt Monomers*

The structure shown in Fig. 5 possesses 3 hcp and 3 fcc non-equivalent hollow sites on the (0001) facet. The adsorption energy of Pt monomer has been calculated for all these sites and found to be within the - 4.77 eV – -5.05 eV range. We find that for all hcp sites $E_{ad}$ is higher (absolute value is lower, binding is weaker) than for any fcc site. In addition, Pt monomers preferably sit on sites surrounded by 2 edge atoms, rather than on those surrounded by only one or none edge atom. Across edge A, the first (1101) available site is a 4-fold hollow site whose adsorption energy, -5.66 eV, is substantially lower than that on hcp sites of the (0001) facet. Across edge B, the first (1101) available site is a 3-fold hollow site, whose adsorption energy is -4.92 eV.

The calculated diffusion barrier $\Delta E$ of the monomer through edge A is found to be highly asymmetric. It is equal to 0.49 eV for diffusion from the fcc site 1 on (0001) to the nearest site in the (1101) facet, and 1.10 eV to diffuse back. Applying the Eq.1 we obtain the rate for the (0001) → (1101) diffusion at the room temperature to be $6 \times 10^3$ s$^{-1}$, which is much lower than that for the diffusion on the facet. The back diffusion rate, however, appears to be negligibly low.

The shortest path for diffusion through the edge B connects the hcp site on the (0001) facet with the three-fold hollow site on the (1101) facet. The barrier for diffusion along this path is found to be 0.28 eV, which is just slightly higher than the barriers for diffusion on the facet. However, the initial state for this process (hcp site) is not an equilibrium state. Its total energy is 0.08 eV higher than that for the monomer adsorbed on the neighboring fcc site. It thus makes sense to consider the diffusion of the Pt monomer through the edge B as a two-step process with diffusion from the fcc to hcp site on (0001) and then from that to the tree-fold site on (1101). This substantially reduces the overall probability of diffusion through the edge B. Moreover, the fcc – hcp barrier is asymmetric (0.20 eV for fcc → hcp and 0.12 eV to diffuse back), which further reduce the diffusion probability. Our results thus suggest that the considered edges are significant obstacles for diffusion of Pt atoms, which may be a factor preventing formation of large Pt islands on the Ru nanoparticles.

*3.2.2. Pt dimers*

The results of previous subsections (3.2.1) suggest that the low diffusion rate of the Pt monomers through the edge between the facets is a factor controlling the size of the Pt islands on the Ru nanoparticles. Since the result for the monomer diffusion through the edge B is not so tractable, we also consider the diffusion of the Pt dimer through that edge.

First, we find $E_{form}/n$ to be higher by 0.12 eV for the dimer than for the monomer, suggesting that dimers would preferably form rather than diffuse as two monomers through the easy edges. As in the case of monomers, dimers prefer to sit on hcp sites on the (0001) facet. On the (0001) facet, when one of the atoms in the dimer comes closer to the edge and its coordination is reduced from 5 to 4, $E_{form}/n$ drops by 0.16 eV, suggesting that there is a higher barrier for Pt dimers to approach the edges to the point at which its atoms become more undercoordinated.

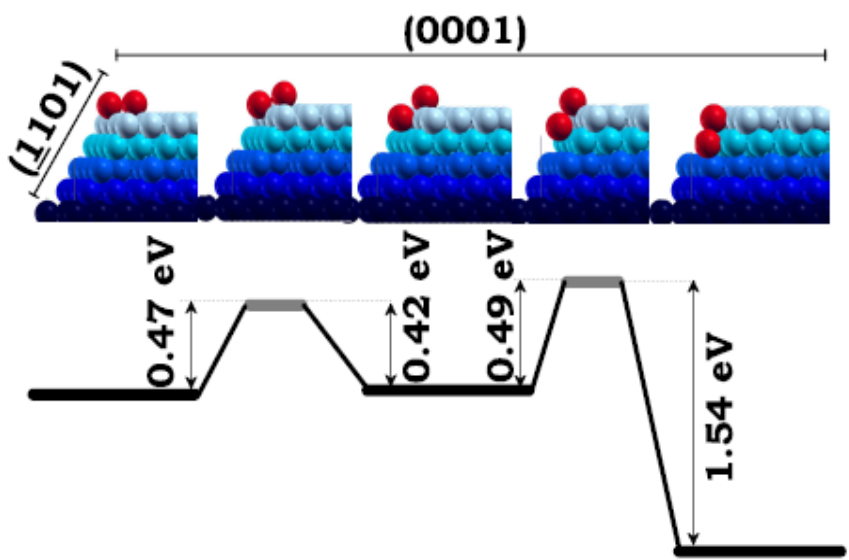

Fig.6

Fig. 6 illustrates the calculated path and energetic for the diffusion of the dimer across the edge B. As seen from the figure, this is a complex process, which comprises two stages with an intermediate energy minimum in which one Pt atom is on the fcc site closest to the edge of (0001), while the other is on the three-fold site in (1101). Due to high barriers the diffusion rate for each step is estimated to be three orders of magnitude lower than those of the monomer suggesting that probability for the Pt dimer to perform the two-step diffusion through the edge B is negligible. As a result of the complex geometric structure of the edge, there could be no diffusion path for the Pt dimer between the facet without significant change of the Pt-Pt and Pt-Ru bond lengths. Such changes increase the total energy of the system and thus result in high diffusion barrier. A simple geometrical analysis suggests that, in the case of trimer or larger islands, this effect are even more pronounced and the rates of their diffusion through the edges are diminishing.

**4. Summary**

As an example of the application of DFT based methods to understand the microscopic processes that control the characteristics of fuel cells catalysts we have presented here a summary of our results which provide a rationale for the observed high CO tolerance of the electrocatalyst based on the Ru nanoparticles with sub-monolayer Pt coverage [6]. We find the energetics of CO adsorbed on the system to be such that CO tends to diffuse from catalytically active Pt onto the Ru substrate. This mechanism requires Pt islands to be small (of few atoms). Meanwhile our further calculations show the propensity of Pt atoms to coalesce and form as large island as possible. This is however true for the flat Ru(0001) surface, while the Ru nanoparticles of interest possesses a number of facets separated by edges. We find that diffusion of Pt atoms and dimers through the edges between Ru facets is prevented by high activation energy barriers. This finding reveals why the nanoscale of the Ru particles is so critical for the process. Pt atoms adsorbed on a particle of such size with sub-monolayer coverage do not move from one facet to the other through the edges. Rather, they form small islands on the facets on which they were adsorbed, resulting in a configuration that is favorable for CO spillover from the active Pt sites, which reduces CO poisoning of this catalyst. The above was one example of many in the literature, which testify to the power of DFT as a reliable tool in computational design of fuel cell catalysts. There are challenges, of course, as DFT results of the type presented here are limited to configurations at 0 K, ultra high vacuum conditions, and to systems containing a few hundred atoms. Attempts are underway to overcome the temperature, pressure, and material gap in computational design of materials for fuel cell catalysts.

**Acknowledgements**
We thank Radislav Adzic for bringing this subject to our attention and for fruitful discussions. The work was supported in part by DOE under grants DE-FG02-07ER15842 and DE-FG02-07ER46354.

**Figure captions**

Fig. 1. Energetics for CO diffusing from the center of Pt islet (c-Pt site) to its edge (e-Pt) and further to Ru substrate (n-Ru and nn-Ru sites). Red, yellow, blue and grey balls represent O, C, Pt and Ru atoms, respectively. Negative and positive numbers correspond to CO adsorption energies and CO diffusion energy barriers, respectively.

Fig. 2. Average formation energy per Pt atom, $E_{form}/n$, as a function of the size, $n$, of the Pt island. Dashed horizontal line marks the $E_{form}/n$ value for one Pt monolayer on Ru(0001).

Fig. 3. (color online) Two configurations of a 7-atom Pt island (blue) on Ru(0001) (grey) showing the detachment of one Pt atom.

Fig. 4. (color online) Two configurations of a 9-atom Pt island (blue) with 2D (left) and 3D (right) structures on Ru(0001) (grey).

Fig. 5. (color online) Model of the edges of a faceted Ru nanoparticle exposing a (0001) facet and two (1101) facets. Different colors distinguish the five layers parallel to the (0001) surface constituting the structure.

Fig. 6. (color online) The upper five panels from left to right illustrate the two-step diffusion of the dimmer (red) across the edge intersecting the (0001) and the (1101) facets (blue). First, third, and fifth upper panels are local minimum energy configurations of the dimmer and the second and forth upper panels are the transition states. The lower panel shows the energetics for iffusion of the dimmer between the (0001) and (1101) facets.